\documentclass{article}

\pdfoutput=1

\usepackage{epsfig}
\usepackage{graphics}
\usepackage{graphicx}
\usepackage[centertags]{amsmath}
\usepackage{amsfonts}
\usepackage{amsthm}
\usepackage{amssymb}
\usepackage{subfigure}
\usepackage{url}

\begin{document}

\title{Complex dynamics emerging in Rule 30 with majority memory}

\author{Genaro J. Mart\'{\i}nez$^1$, Andrew Adamatzky$^1$ \\ Ramon Alonso-Sanz$^{1,2}$ and J.C. Seck-Touh-Mora$^3$}


\maketitle

\begin{centering}
$^1$ Department of Computer Science, University of the West of England, Bristol, United Kingdom. \\
\url{{genaro.martinez, andrew.adamatzky}@uwe.ac.uk} \\

$^2$ ETSI Agr\'onomos, Polytechnic University of Madrid, Madrid, Spain. \\
\url{ramon.alonso@upm.es} \\

$^3$ Centro de Investigaci\'on Avanzada en Ingenier\'{\i}a Industrial, Universidad Aut\'onoma del Estado de Hidalgo Pachuca, Hidalgo, M\'exico. \\
\url{jseck@uaeh.edu.mx} \\
\end{centering}

\begin{abstract}
\noindent
In cellular automata with memory, the unchanged maps of the conventional cellular automata are  applied to cells  endowed with memory  of their past states in some specified interval. We implement Rule 30 automata with a majority memory and show that using the memory function we can transform quasi-chaotic dynamics of classical Rule 30 into domains of travelling structures with predictable behaviour. We analyse morphological complexity of the automata and classify dynamics of gliders (particle, self-localizations) in memory-enriched Rule 30. We provide formal ways of encoding and classifying glider dynamics using de Bruijn diagrams, soliton reactions and quasi-chemical representations. 

\vspace{0.5cm}

\noindent
\textbf{Keywords:} Rule 30, memory, elementary cellular automata, chaos, complexity dynamics, gliders, emergency.
\end{abstract}

\section{Introduction}
Elementary cellular automaton (CA) is a one-dimensional array of finite automata, each automaton takes two states and updates
its state in discrete time depending on its own state and states of its two closest neighbours, all cell updates their state synchronously. A general classification of elementary CA was introduced in \cite{kn:Wolf94}:

\begin{quote}
\begin{enumerate}
\item[] Class I. CA evolving to a homogeneous state.
\item[] Class II. CA evolving periodically.
\item[] Class III. CA evolving chaotically.
\item[] Class IV. Include all previous cases, also know as class of complex rules.
\end{enumerate}
\end{quote}

In this classification Class IV is of particular interest because rules of the class exhibit non-trivial behaviour with rich and diverse patterns, e.g. Rule 54 CA~\cite{kn:MAM06,kn:MAM08}.

\section{Basic notation}

\subsection{One-dimensional cellular automata}
One-dimensional CA are represented by an infinite array of {\it cells} $x_i$ where $i \in \mathbb{Z}$ and each $x$ takes a value from a finite alphabet $\Sigma$. Thus, a sequence of cells \{$x_i$\} of finite length $n$ represents a string or {\it global configuration} $c$ on $\Sigma$. This way, the set of finite configurations will be represented as $\Sigma^n$. An evolution is represented by a sequence of configurations $\{c_i\}$ given by the mapping $\Phi:\Sigma^n \rightarrow \Sigma^n$; thus their global relation is provided as follows

\begin{equation}
\Phi(c^t) \rightarrow c^{t+1}
\label{globalFunction}
\end{equation}

\noindent where $t$ is time and every global state of $c$ is defined by a sequence of cell states. Also the cell states in configuration $c^t$ are updated at the next configuration $c^{t+1}$ simultaneously by a local function $\varphi$ as follow

\begin{equation}
\varphi(x_{i-r}^t, \ldots, x_{i}^t, \ldots, x_{i+r}^t) \rightarrow x_i^{t+1}.
\label{localFunction}
\end{equation}

Wolfram represents one-dimensional CA with two parameters $(k,r)$. Where $k = |\Sigma|$ is the number of states, and $r$ is radius of neighboourhood. Elementary CA are defined by parameters $(k=2,r=1)$. There are $\Sigma^n$ different neighborhoods (where $n=2r+1$) and $k^{k^n}$ different evolution rules.

In computer experiments we were using automata with periodic boundary conditions.

\subsection{Cellular automata with memory}

Conventional cellular automata are ahistoric (memoryless): i.e., the new state of a cell depends on the neighborhood configuration solely at the preceding time step of $\varphi$ (eq.~\ref{localFunction}).

CA with {\it memory} considers an extension to the standard framework of CA by implementing memory capabilities in cells $x_i$ from its own history.

Thus to implement memory we incorporate a memory function $\phi$, as follow:

\begin{equation}
\phi (x^{t-\tau}_{i}, \ldots, x^{t-1}_{i}, x^{t}_{i}) \rightarrow s_{i}
\end{equation}

\noindent such that $\tau < t$ determines the degree of memory backwards and each cell $s_{i} \in \Sigma$ being a state function of the series of states of the cell $x_i$ with memory up to time-step. Finally to execute the evolution we apply the original rule as:

$$
\varphi(\ldots, s^{t}_{i-1}, s^{t}_{i}, s^{t}_{i+1}, \ldots) \rightarrow x^{t+1}_i.
$$

Thus in CA with memory,  while the mappings $\varphi$ remain unaltered, historic memory of all past iterations is retained by featuring each cell by a summary of its past states from $\phi$. Therefore cells {\it canalize} memory to the map $\varphi$.

As an example, we can see a case of memory $\phi$. The {\it majority memory} is defined as follow:

\begin{equation}
\phi_{maj} \rightarrow s_{i}
\end{equation}

\noindent where in case of a tie given by $\Sigma_1 = \Sigma_0$ from $\phi$ then we will take the last value $x_i$. So $\phi_{maj}$ function represents the classic majority function \cite{kn:Mins67} on the cells $(x^{t-\tau}_{i}, \ldots, x^{t-1}_{i}, x^{t}_{i})$ and define a temporal ring before to get finally the next global configuration $c$.

\begin{figure}[th]
\centerline{\includegraphics[width=4.8in]{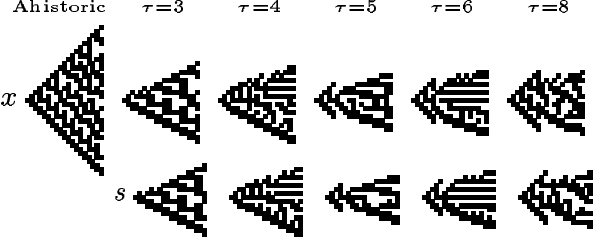}}
\caption{The effect of majority memory of increasing depth on Rule 30 starting from a single site live cell.}
\label{memoryexample}
\end{figure}

Majority memory exerts a general inertial effect \cite{kn:Alo08}. This effect, when starting from a single site live cell notably restrains the dynamics, as it is illustrated on Rule 30 in fig.~\ref{memoryexample}. This figure shows the spatio-temporal patterns of both the $current$ $x$ state values and that of the underlying $s$ ones.

\section{Elementary CA Rule 30}
Rule 30 was initially studied by Wolfram in \cite{kn:Wolf94} because of its chaotic global behavior; looking for a number random generator. Rule 30 is an elementary CA evolving in one dimension of order $(2,1)$. An interesting property is that Rule 30 has a surjective relation and thus do not have {\it Garden of Eden} configurations \cite{kn:AC70}. In this way, any configuration has always at least one predecessor. 

The local rule $\varphi$ corresponding to Rule 30 is following:

\[
\varphi_{R30} = \left\{
	\begin{array}{lcl}
		1 & \mbox{if} & 100, 011, 010, 001 \\
		0 & \mbox{if} & 111, 110, 101, 000
	\end{array} \right. .
\]

Generally speaking, Rule 30 displays a typical chaotic global behavior --- Class III in Wolfram's classification. An interesting study on Rule 30 showing a local nested structure that repeat periodically looking invertible properties \cite{kn:Row06}.

So, initially $\varphi_{R30}$ has a 50\% of probability of states zero or one, and consequently each state appears with the same frequency.

\begin{figure}[th]
\centerline{\includegraphics[width=4.8in]{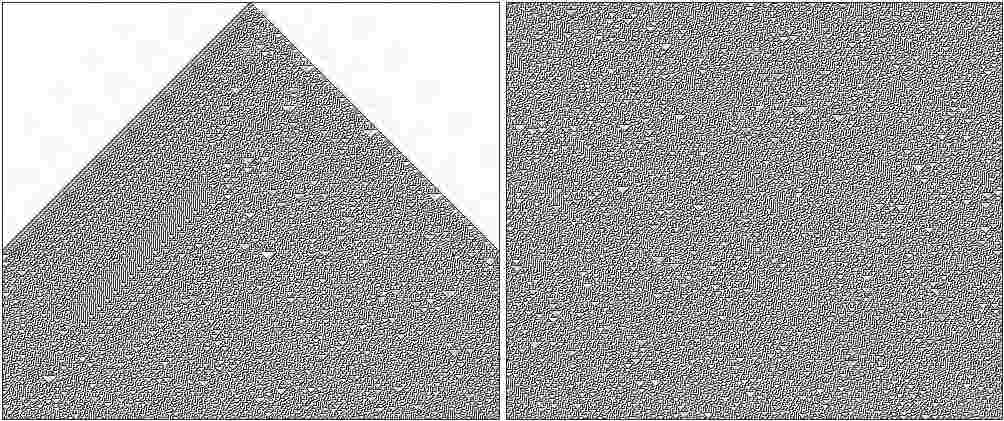}}
\caption{Typical behavior of Rule 30 where a single cell in state 1 leads to chaotic state (left). Second configuration (right) shows automaton behavior from a random initial condition with initial density of 50\% for each state. Both automata evolve on a ring of 497 cells (with periodic boundary property) to 417 generations. White cells represent state 0 and dark cells the state 1.}
\label{R30evol}
\end{figure}

Also the evolution rule presents the following feature: if an initial configuration is covered only of state one then the configuration always evolves into one but if this is empty or filled with state one then this always evolves to state zero. Fig.~\ref{R30evol} shows such two cases of typical evolutions in Rule 30.

\subsection{De Bruijn and subset diagrams in Rule 30}
Given a finite sequence $w \in \Sigma^m$; such that $w=w_1, \ldots, w_m$; let $\alpha(w)=w_1$, $\beta(w)=w_2, \ldots,w_m$, and $\psi(w)=w_1, \ldots,w_{m-1}$. With these elements, we can specify a labeled directed graph known as de Bruijn diagram $\mathcal{B}=\{N;E\}$ associated with the evolution rule of the CA. The nodes of $\mathcal{B}$ are defined by $N=\Sigma^{2r}$ and the set of directed edges $E \subseteq \Sigma^{2r} \times \Sigma^{2r}$ is defined as follows:

\begin{equation}
E=\{(v,w)\, \mid \, v,w \in N,\; \beta(v)=\psi(w) \}.
\end{equation}

\begin{figure}[th]
\centerline{\includegraphics[width=1.7in]{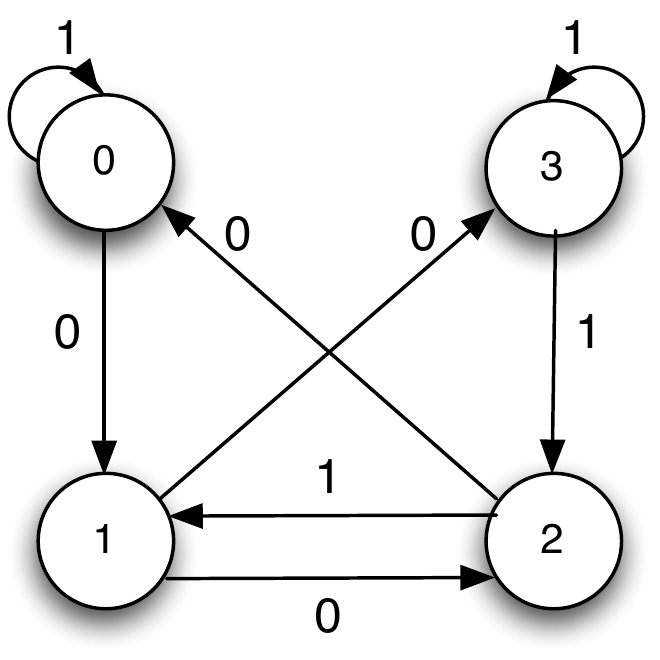}}
\caption{De Bruijn diagram for the elementary CA Rule 30.}
\label{dB30_01}
\end{figure}

For every directed edge $(v,w) \in E$, let $\eta(v,w) = aw \in \Sigma^{2r+1}$ where $a=\alpha(v)$; that is, $\eta(v,w)$ is a neighborhood of the automaton. In this way, the edge $(v,w)$ is {\it labeled} by $\varphi \circ \eta(v,w)$; hence every labeled path in $\mathcal{B}$ represents the evolution of the corresponding sequence specified by its nodes. Since each $w \in N$ can be described by a number base $k$ of length $2r$, every node in $\mathcal{B}$ can be enumerated by a unique element in $\mathbb{Z}_{k^{2r}}$, which is useful to simplify the diagram. The de Bruijn diagram associated to Rule 30 is depicted in fig. \ref{dB30_01}; where black edges indicate the neighborhoods evolving into $0$ and the ones evolving into $1$ are shown by gray edges.\footnote{De Bruijn and subset diagrams were calculated using NXLCAU21 designed by McIntosh; available from \url{http://delta.cs.cinvestav.mx/~mcintosh}}

\begin{figure}[th]
\centerline{\includegraphics[width=4.8in]{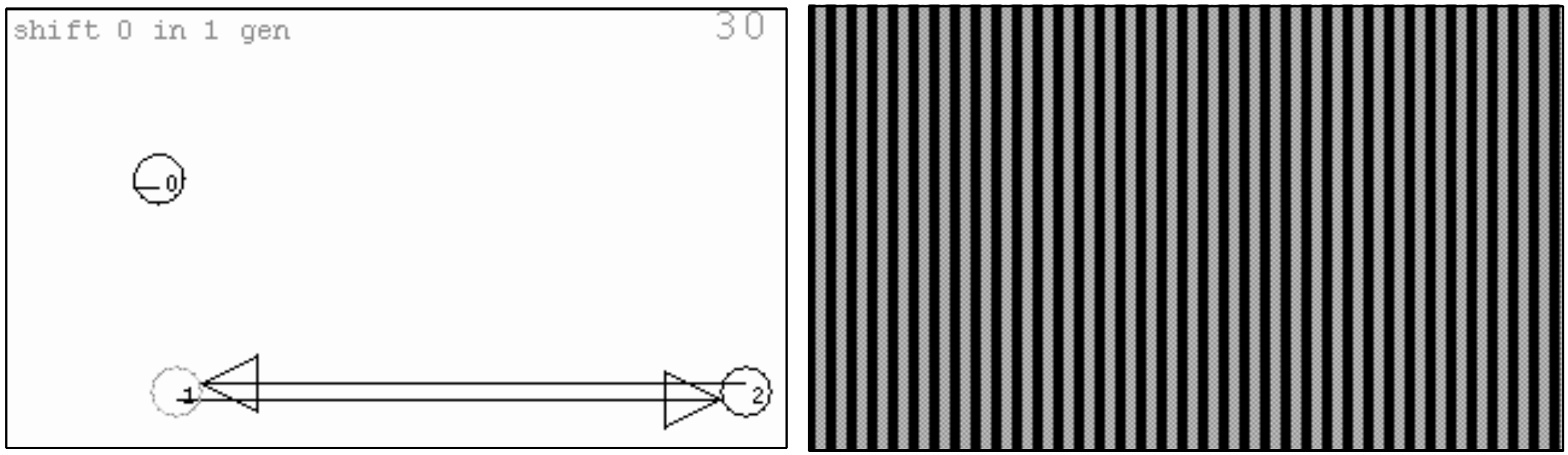}}
\caption{Cycles in the de Bruijn diagram and the corresponding periodic evolution for cycle $(1,2)$.}
\label{dB30_02}
\end{figure}

Fig.~\ref{dB30_01} shows that there are four neighborhoods evolving into $0$ and four into $1$; so each state has the same probability to appear during the evolution of the automaton; indicating the possibility of the automaton is surjective, i.e. 
there are no Garden of Eden configurations. Classical analysis in graph theory has been applied over de Bruijn diagrams for studying topics such as reversibility \cite{kn:Seck05}: cycles in the diagram indicate periodic elements in the evolution of the automaton if the label of the cycle corresponds to the sequence defined by its nodes, in periodic boundary conditions. 
The cycles in the de Bruijn diagram from fig.~\ref{dB30_01} are presented in fig.~\ref{dB30_02}.

The largest cycle in fig. \ref{dB30_02} indicates that the undefined repetition of sequence $w_b = 10$ establishes a periodic structure without displacement in one generation during the evolution of Rule 30 we will than $w_b$ is the {\it filter}\footnote{A {\it filter} is a periodic sequence that exist alone or in blocks into of the evolution, thus a suppression of such string produce a new view.} in Rule 30; in this case, the filter reported in the original Rule 30 and with memory too. Thus we can see how a de Bruijn diagram can recognize any periodic structures in CA \cite{kn:MMS08,kn:MAM08}.

De Bruijn diagram is nondeterministic in the sense that a given node may have several output edges with the same label; a classical approach to analyse the diagram would be to construct the subset (or power) diagram in order to obtain a deterministic version for the de Bruijn diagram in the evolution rule \cite{kn:Mc,kn:Voor96}.

\begin{figure}[th]
\centerline{\includegraphics[width=3in]{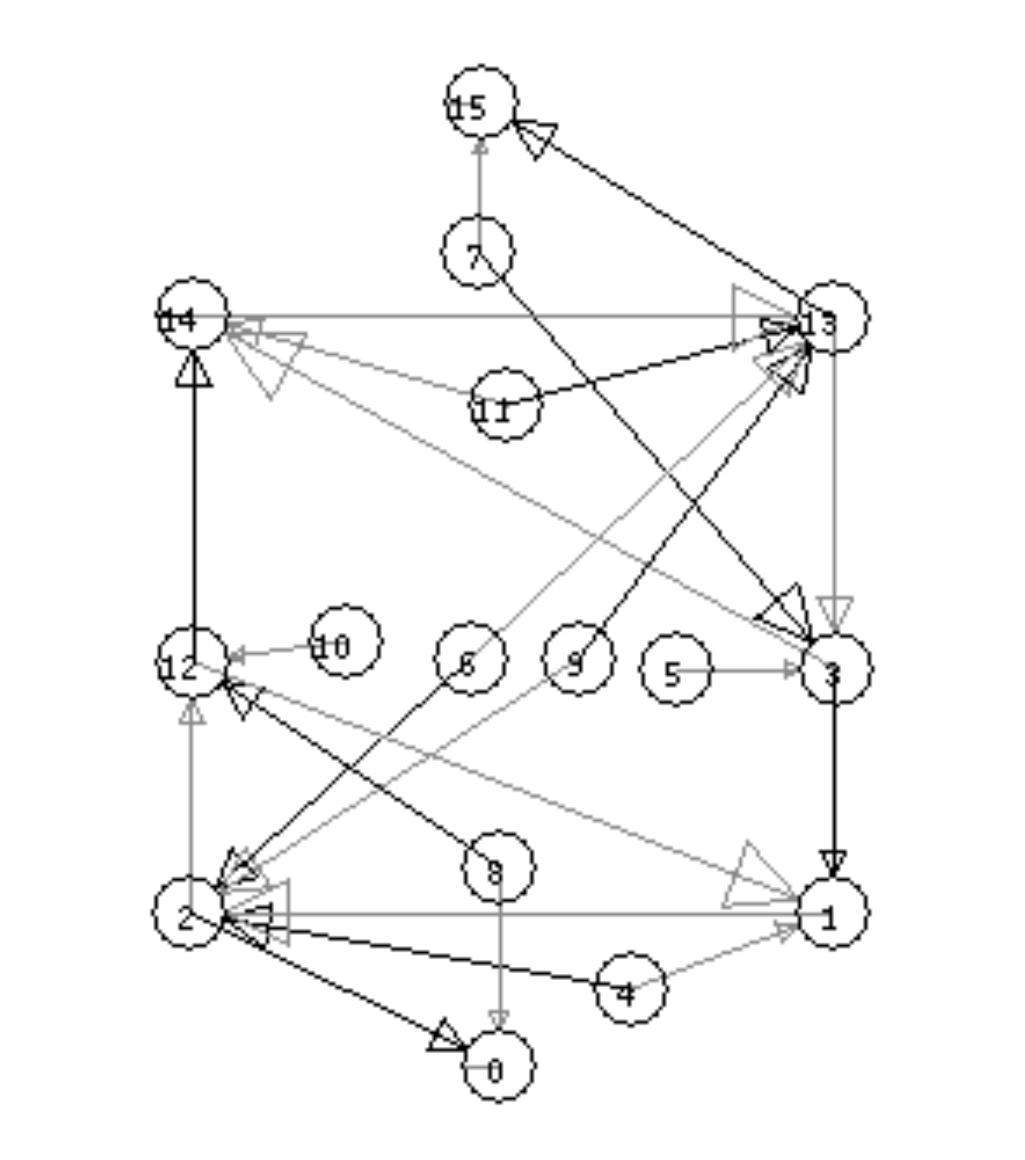}}
\caption{Subset diagram for Rule  30.}
\label{ss30_01b}
\end{figure}

The subset diagram is defined as $\mathcal{S}=\{\mathcal{P},\mathcal{Q}\}$ where $\mathcal{P}=\{P \mid P \subseteq \Sigma \cup \emptyset \}$ is the set of nodes of $\mathcal{S}$ and the directed edges are defined by $\mathcal{Q} \subset \mathcal{P} \times \mathcal{P}$ where for $P_1, P_2 \in \mathcal{P}$ there is a directed edge $(P_1,P_2)$ labeled by $a \in S$ in $\mathcal{S}$ iff 
$P_2$ is the maximum subset such that for every $c \in P_2$ there exists $b \in P_1$ such that $\varphi(b,c)=a$.

The inclusion of the empty set assures that every edge has a well-defined ending node. For a CA with $k$ states, it is fulfilled that $|\mathcal{P}|=2^{k^{2r}}$ which implies an exponential growth in the number of nodes in $\mathcal{S}$ when more states are considered. Every $P \in \mathcal{P}$ can be identified by a binary number showing the states belonging to this subset; that is, taking the states as an ordered list, the ones in $P$ can be signed by a $1$ and the others by $0$; conforming a unique binary sequence identifying the subset. The decimal value of this binary number can be taken to get a shorter representation; where the empty set has a decimal number $0$ and the full subset $p=\Sigma$ has the number $2^{k^{2r}}-1$. The a subset diagram corresponding to Rule 30 is shown in fig.~\ref{ss30_01b}.

In fig. \ref{ss30_01b}, the subset diagram has no path starting from the full subset (node $15$) going to the empty one (node $0$).
This means that every sequence can be produced by the evolution of the automaton and there are no Garden-of-Eden sequences. Thus the automaton is surjective. The subset diagram can be also used as a deterministic automaton for calculating ancestors of any desired sequences~\cite{kn:Seck04} by recognizing the regular expressions which may be generated by the corresponding automaton. Some of these expressions would be able to represent interesting structures as gliders \cite{kn:MMS08b}; however more effort is needed in order to get a straightforward detection of such constructions in the diagram.

Finally, such diagrams helps to get periodic strings than eventually represent a general filter $w_b$ working at the original Rule 30 and Rule 30 with memory (next section). Also, we will take advantage of these results to find gliders from these strings.

\section{Majority memory helps to discover complex dynamics in Rule 30}
This section reports how the majority memory $\phi$ helps to discovers complex dynamics in elementary CA by experimentation, see an introduction to elementary CA with memory in~\cite{kn:AM03,kn:AM05,kn:Alo06}.

\begin{figure}
\centerline{\includegraphics[width=4.8in]{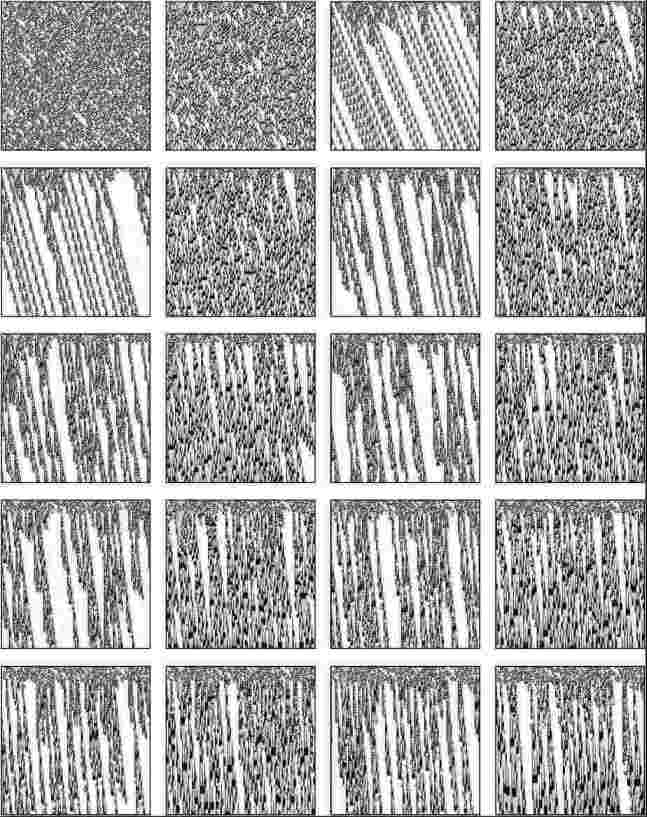}}
\caption{Complex dynamics emerging in Rule 30 with majority memory $\phi_{maj}$ from a range of values from $\tau=3$ to $\tau=21$ although the first evolution shows the original function. Evolutions were calculated on a ring of 104 cells in 104 generations with a random initial density of 50\% where the initial condition is the same in all cases. Also the filter $w_b$ was applied to see clearly the structures.}
\label{mMemory}
\end{figure}

In this way, fig.~\ref{mMemory} displays different scenarios where the majority memory works on Rule 30 and therefore extracting complex dynamics on $\phi_{maj}$. We should read the evolutions from left to right and up to down. Also in this case all evolutions use the same random initial density and the filter $w_b$ was utilized in all evolutions (including the original Rule 30 evolution).

Thus the first evolution is the original function Rule 30, i.e, without majority memory in fig~\ref{mMemory}. From its original evolution we can see gaps that the filter can clean. Traditionally was difficult distinguish such filter in Rule 30 but when $\phi_{maj}$ was applied then its presence was more evident. A general technique to get filters was developed by Wuensche in \cite{kn:Wue99}.

Initially, even values of $\tau$ seems extract gliders more quickly and odd values fight to reach an order. Eventually the majority memory will converge to one stability in $\Phi$ while $\tau$ increases.

The first snapshot calculating $\phi_{maj}$ with $\tau = 3$ was the second evolution in fig~\ref{mMemory}. Nevertheless, here is not very clear how the memory could induce another behavior because its global behavior is similar at the original but with small changes.

On the other hand, the third evolution with $\tau=4$ extracts periodic patterns emerging in Rule 30 with memory. The evolution maybe does not display impressive gliders but it already allows to pick up more diversity in mobile localizations on lattices of $100 \times 100$ cells. Thus we have enumerated and ordered values of $\tau$ from fig.~\ref{mMemory} based on their space-time dynamics they are responsible for.

\begin{quote}
\begin{enumerate}
\item[] {\it Chaotic global behavior} $\tau=0,3,5,7,9,11,13,15,17,19,21$.
\item[] {\it Periodic patterns} $\tau=4,6,8,10,12,14,16,18,19,20,21$.
\item[] {\it Collision patterns} $\tau=6,8,10,12$.
\end{enumerate}
\end{quote}

\subsection{Morphological complexity in Rule 30 with memory}
Looking a way to explore global complex dynamics in Rule 30 and itself with memory $\phi_{maj}$. In this section we will explore some techniques.

We evaluate morphological complexity of CA studied using morphological richness approach \cite{kn:AH98}. We calculate statistical morphological richness $\mu$ as follows. Given space-time configuration of a one-dimensional CA, for each site of the configuration we extract its $3 \times 3$ cell space-time neighborhood state and build distribution of the space-time neighborhood states over extended period of the automaton's development time.

\begin{figure}
\centering
\subfigure[]{\includegraphics[width=0.49\textwidth]{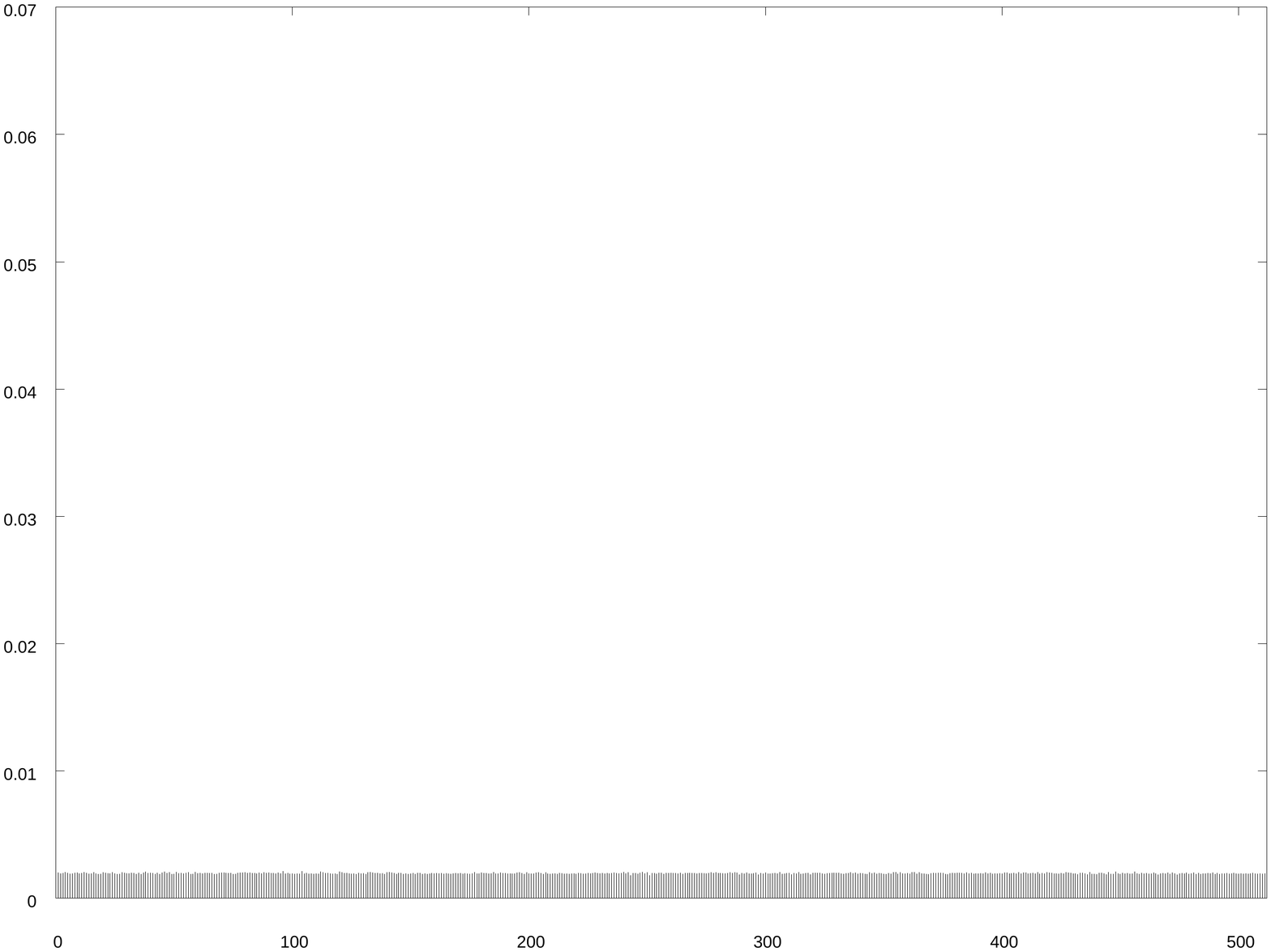}}
\subfigure[]{\includegraphics[width=0.49\textwidth]{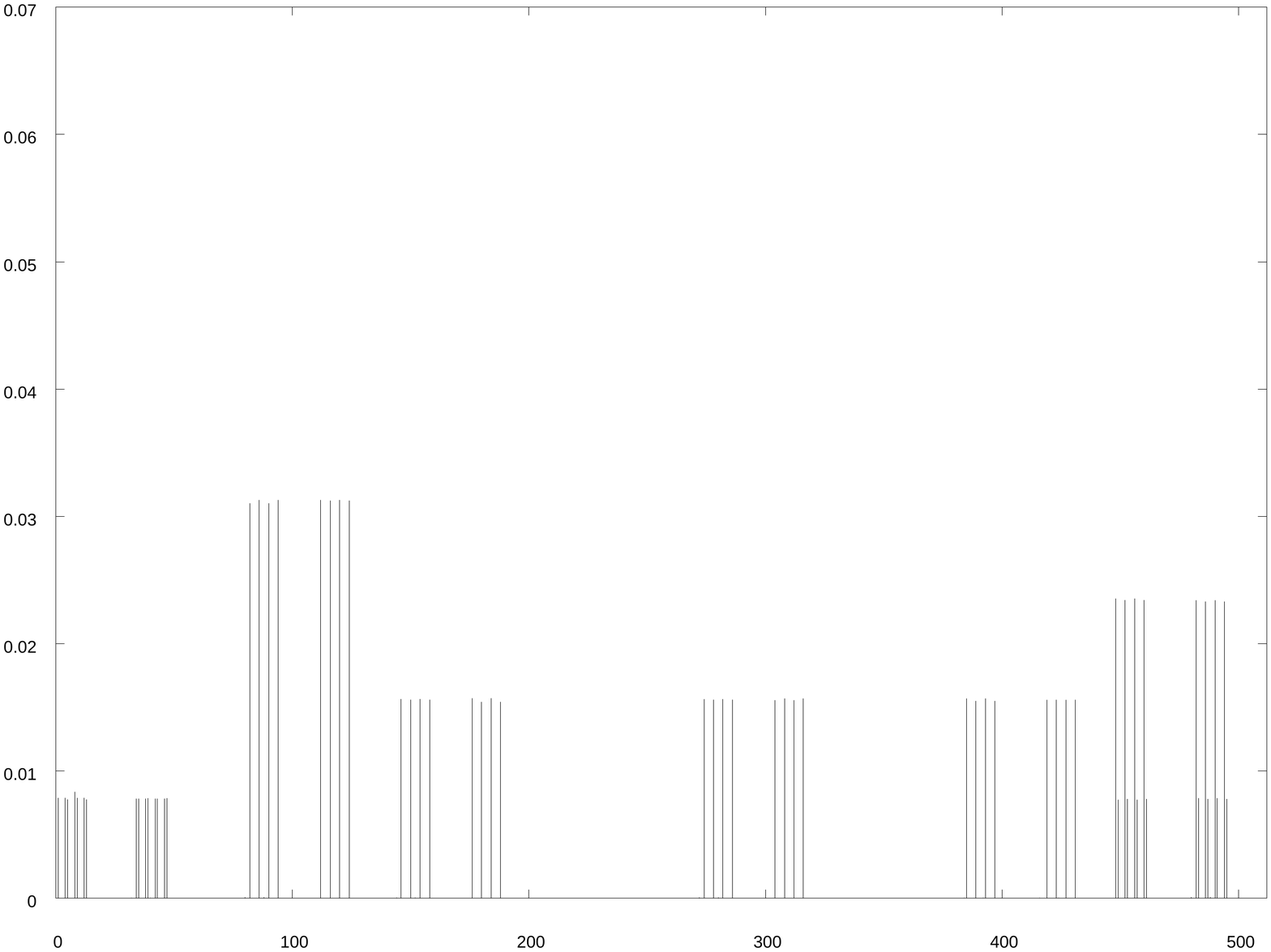}}
\subfigure[]{\includegraphics[width=0.49\textwidth]{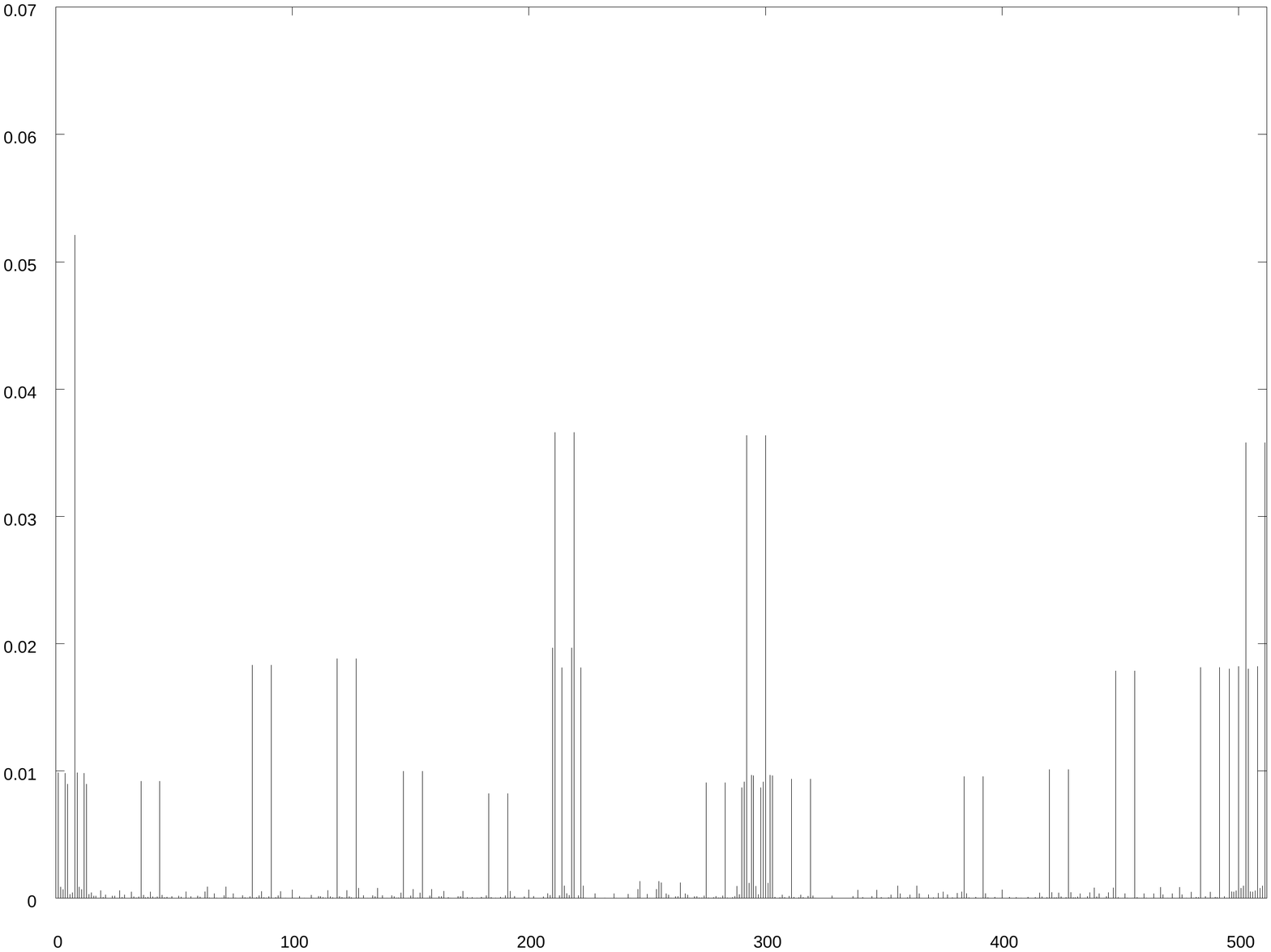}}
\subfigure[]{\includegraphics[width=0.49\textwidth]{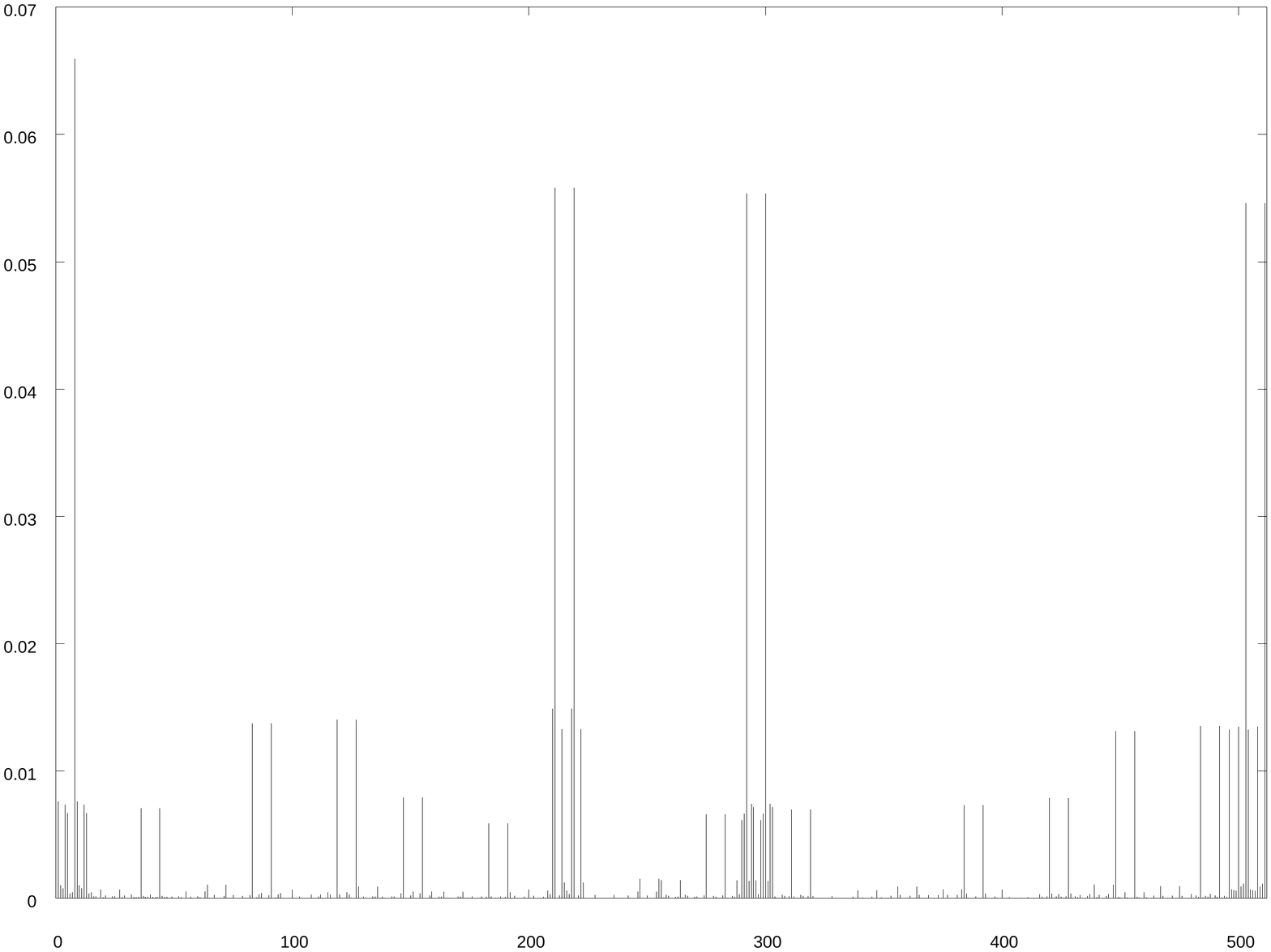}}
\subfigure[]{\includegraphics[width=0.49\textwidth]{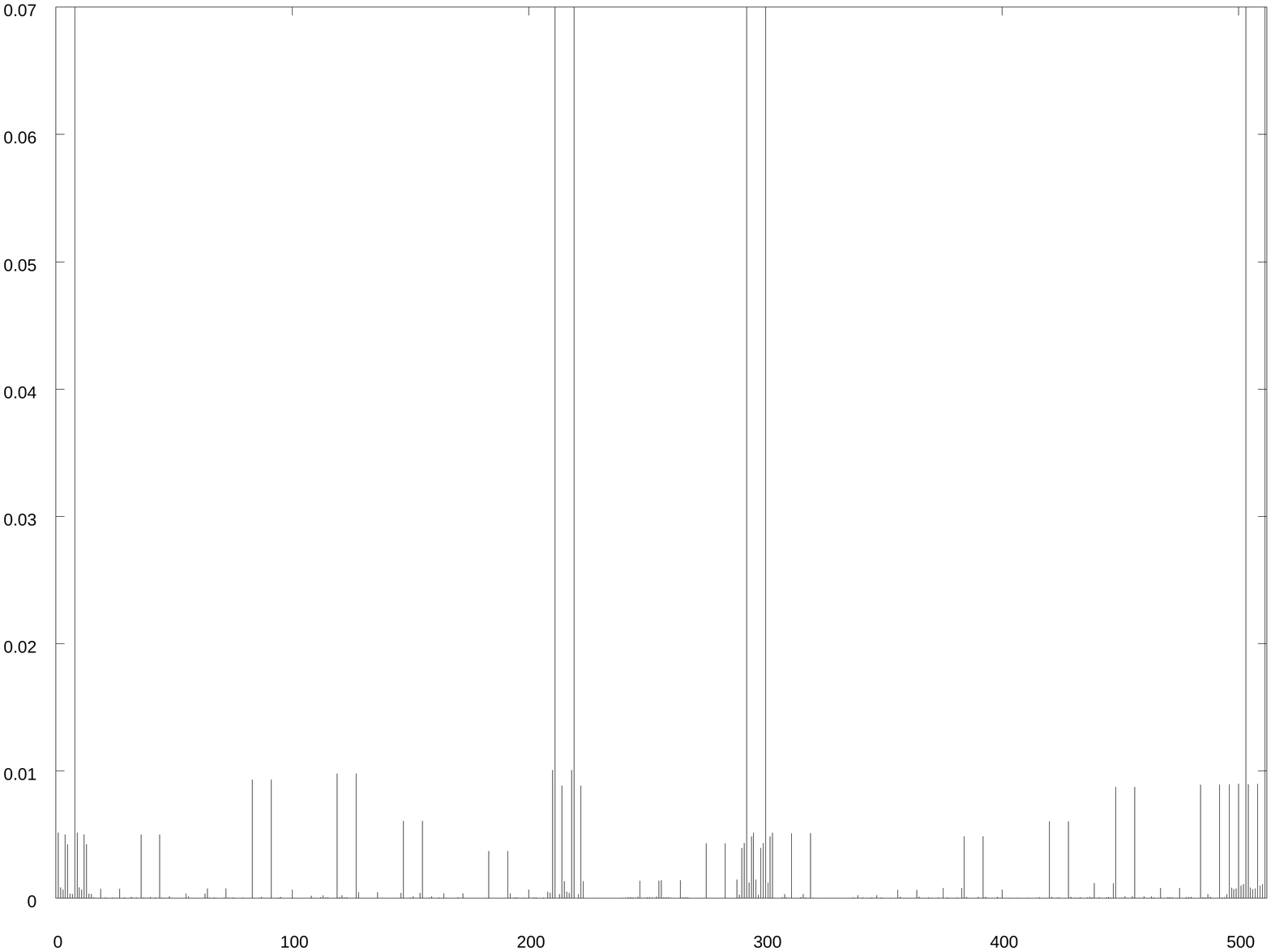}}
\subfigure[]{\includegraphics[width=0.49\textwidth]{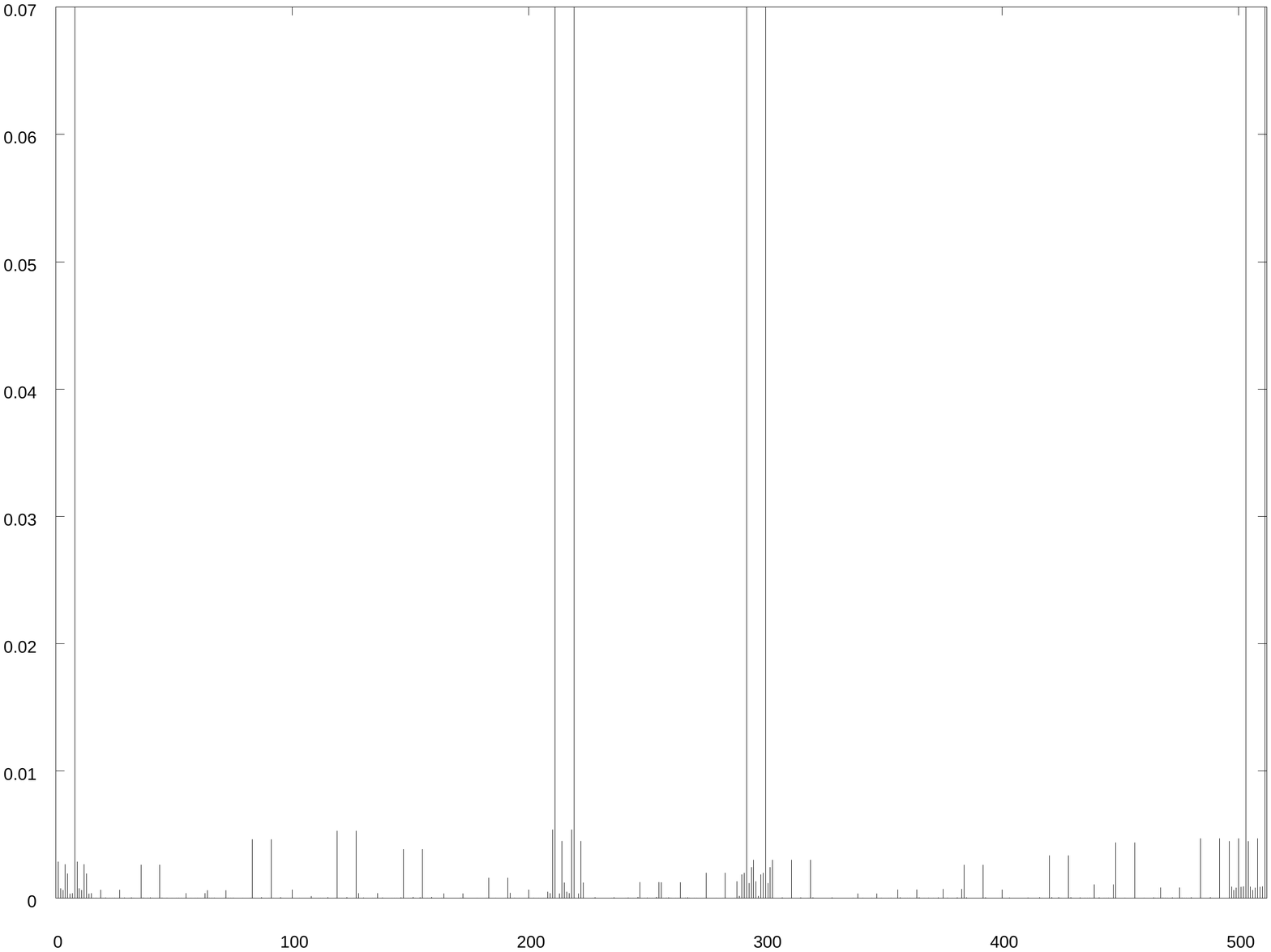}}
\caption{Morphological richness. CA length 1500 cells, running time 5000 steps. (a)~random update of cell states, (b)~Rule 30 without memory, (c)--(f) with memory, (c)~$\tau=3$, (d)~$\tau=5$, (e)~$\tau=10$, (f)~$\tau=21$.}
\label{2Dmorphology}
\end{figure}

Examples of morphological richness $\mu$ are shown in fig.~\ref{2Dmorphology}. A control case, when next state of a cell is calculated at random the distribution (fig.~\ref{2Dmorphology}a) of space-time neighborhood states is uniform. Two-dimensional random configurations are morphologically rich. Morphology of memoryless, classical, Rule 30 automaton is characterized by few peaks in local domains distributions, where several space-time templates dominate in the global space-time configuration (fig.~\ref{2Dmorphology}b). The statistical morphological richness $\mu$ decreases.

Incorporation of a memory in the cell-state transition rules leads to erosion of the distribution (fig.~\ref{2Dmorphology}c) and thus slight increases in $\mu$. With increase of memory depth shape of the morphological distribution changes just slightly, up to minor variations in heights of major peaks (fig.~\ref{2Dmorphology}d--f).

Number $\rho$ of $3 \times 3$ blocks (of states 0 and 1) that never appear in space-time configurations of a cellular automaton can be used as a express estimate of the nominal morphological richness, the less $\rho$ the more rich is a nominally configuration.

The difference between statistical $\mu$ and nominal $\rho$ measures of morphological richness is that $\mu$ allows to pick up most common configurations of local domains, while $\rho$ just shows how many block of $ 3 \times 3$ states appeared in the automaton evolution at least once.

For the case of random update of cell states, all blocks are present in the space-time configuration, $\rho=0$.

Memoryless automata governed by Rule 30 have $\rho=434$ (total number of possible blocks is 512). When memory is incorporated in the cell-state transition function, first richness decreases, e.g. for $\tau=1$, we have $\rho=448$. Then we observe consistent increase in complexity. Thus Rule 30 with small-depth memory ($\tau=2$) $\rho=140$, which drastically decrease to $\rho=68$ for $\tau=3$. The richness is stabilized, or rather oscillate around values $\rho=20-40$ with further increase of memory.

In summary, we found that majority memory increases nominal complexity of cellular automata but decreases statistical complexity.

\subsection{Gliders in Rule 30 with memory $\tau=8$}

Most frequently the complex dynamics related in elementary CA is related to gliders, glider guns, and non-trivial reactions between localizations, for example, Rule 110 or 54 \cite{kn:MMS06,kn:MAM06}. The relevance to look for such things is the study of universality \cite{kn:Cook04,kn:Wolf02}. Another additional subject was its representation by regular expressions and other tools \cite{kn:MMS08,kn:MAM08}.

\begin{figure}[th]
\centerline{\includegraphics[width=4.8in]{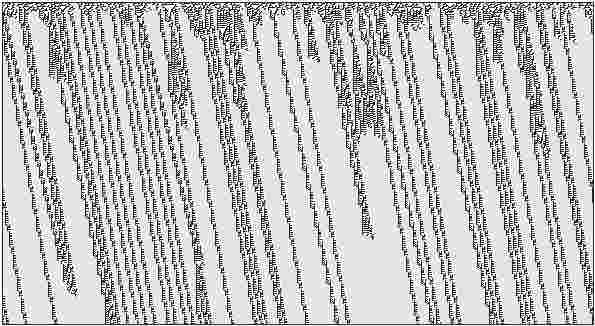}}
\caption{Gliders emerging in Rule 30 with $\phi_{maj}$ and $\tau=8$. This evolution shows how some kinds of gliders arise and still interact from random initial conditions. The evolution was calculated on a ring of 590 cells to 320 generations, and an initial density of 50\%.}
\label{m8-memory}
\end{figure}

Among the sets of complex dynamics in Rule 30 determined by $\tau$ (showed in fig.~\ref{mMemory}), we have chosen the memory $\phi_{maj}$ with $\tau=8$. In this way, fig~\ref{m8-memory} illustrates an ample evolution space of its global dynamics.

Of course, these gliders maybe are not as impressive as others from well-known complex rules as Rule 110, Rule 54 or other one dimensional rules \cite{kn:BNR91,kn:MMS06,kn:MAM06,kn:Wolf94,kn:Wue94}. However it is interesting how $\phi_{maj}$ is able to open complex patterns from chaotic rules.

Nevertheless, even that Rule 30 does not offer an ample range of complex dynamics, it is useful to describe gliders and collisions. So, we shall illustrate how a chaotic CA can be discomposed as a complex system.

\begin{figure}[th]
\centerline{\includegraphics[width=3.8in]{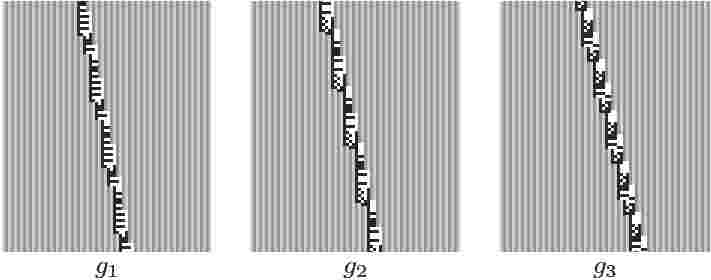}}
\caption{Set of gliders $G_{R30m}$ with memory $\phi_{maj}$ and $\tau=8$.}
\label{gliders}
\end{figure}

Thus we classify the family of gliders and enumerate some properties. Then fig.~\ref{gliders} displays the family of gliders $G_{R30m} = \{g_1, g_2, g_3\}$. Although as was hoped, an immediate consequence was that gliders in CA with memory have longer periods.

\begin{table}[th]
\centering
\begin{tabular}{|c|c|c|}
\hline
structure & $v_{g}$ & lineal volume \\
\hline \hline
$w_b$ & $0/c$ = 0 & 2 \\
\hline
$g_{1}$ & $2/11 \approx 0.1818$ & 5 \\
\hline
$g_{2}$ & $4/19 \approx 0.2105$ & 7 \\
\hline
$g_{3}$ & $4/17 \approx 0.2352$ & 6 \\
\hline
\end{tabular}
\caption{Properties of gliders $G_{R30m}$ with memory $\phi_{maj}$ and $\tau=8$.}
\label{tablaGlidersR30}
\end{table}

Tab.~\ref{tablaGlidersR30} summarizes the basic properties of each glider. Particularity all gliders in this domain have a constant displacement of four cells to the right and none glider with speed zero was founded, and yet it was more complicated to find a glider gun in this domain. Nevertheless, some interesting reactions were originated from $G_{R30m}$.

Structure $w_b$ does not have a displacement and also it is neither a glider. This pattern is the periodic background in Rule 30 and still represent the filter in Rule 30. It was really hard to detect the existence of a periodic background evolving the original rule; but when $\phi_{maj}$ was applied, a periodic pattern began to emerge inherited from $\varphi_{R30}$. Finally this filter was confirmed with its respective de Bruijn and cycle diagrams (see section 3.1).

\subsection{Reactions between gliders from $G_{R30m}$}
Let us demonstrate simple examples of collisions between gliders. Codes for all gliders, necessary to generate the whole 
set of binary collisions, are presented in appendixes~A and B.

Fig.~\ref{cylce-g3g2g1} shows how a stream of $g_1$ gliders is deleted from a reaction cycle:

$$
g_3 + g_1 \rightarrow g_2 \mbox{ and } g_2 + g_1 \rightarrow g_3.
$$

\begin{figure}[th]
\centerline{\includegraphics[width=4.75in]{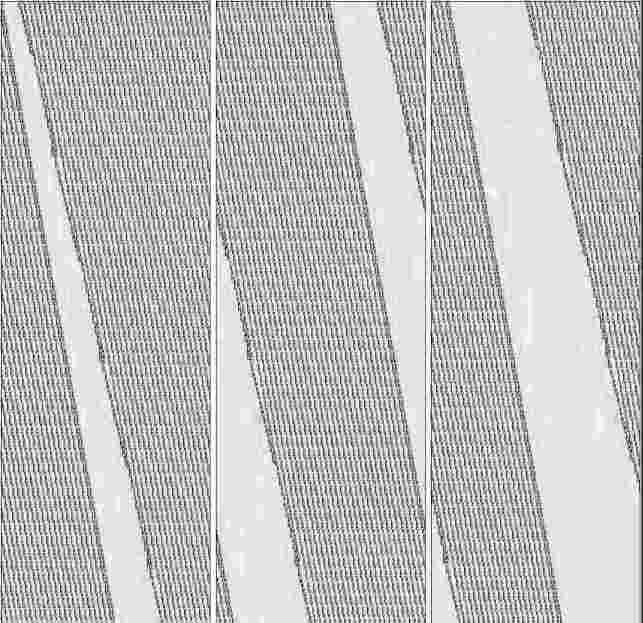}}
\caption{Deleting streams of $g_1$ gliders initialized with a single $g_3$ glider.}
\label{cylce-g3g2g1}
\end{figure}

To obtain such cycle by glider reaction we can code an unlimited initial condition as $...g_3...g_1....g_1....g_1....$, that can be reduced as $...g_3...g_1(....g_1)^*$ (where a dot represent a copy of $w_b$). Finally the evolution produces the cycle above, which is produced with nine periods of $g_3$ and eight periods of $g_2$. Thus each column presents 1135, 2270 and 3405 generations respectively.\footnote{The glider reactions were produced using OSXLCAU21 system available at~\url{http://uncomp.uwe.ac.uk/genaro/OSXCASystems.html}}. However in this case the codification of gliders is affected because could not natural as was done in other cases \cite{kn:MMS06,kn:MMS08,kn:MAM08}.

Since the better way to preserve $\varphi_{R30}$ and $\phi_{maj}$ to code gliders as ``natural,'' we will consider the codification from its original initial condition (see appendix B where some strings are defined to get gliders with memory from its original function).

\begin{figure}[th]
\centerline{\includegraphics[width=4.75in]{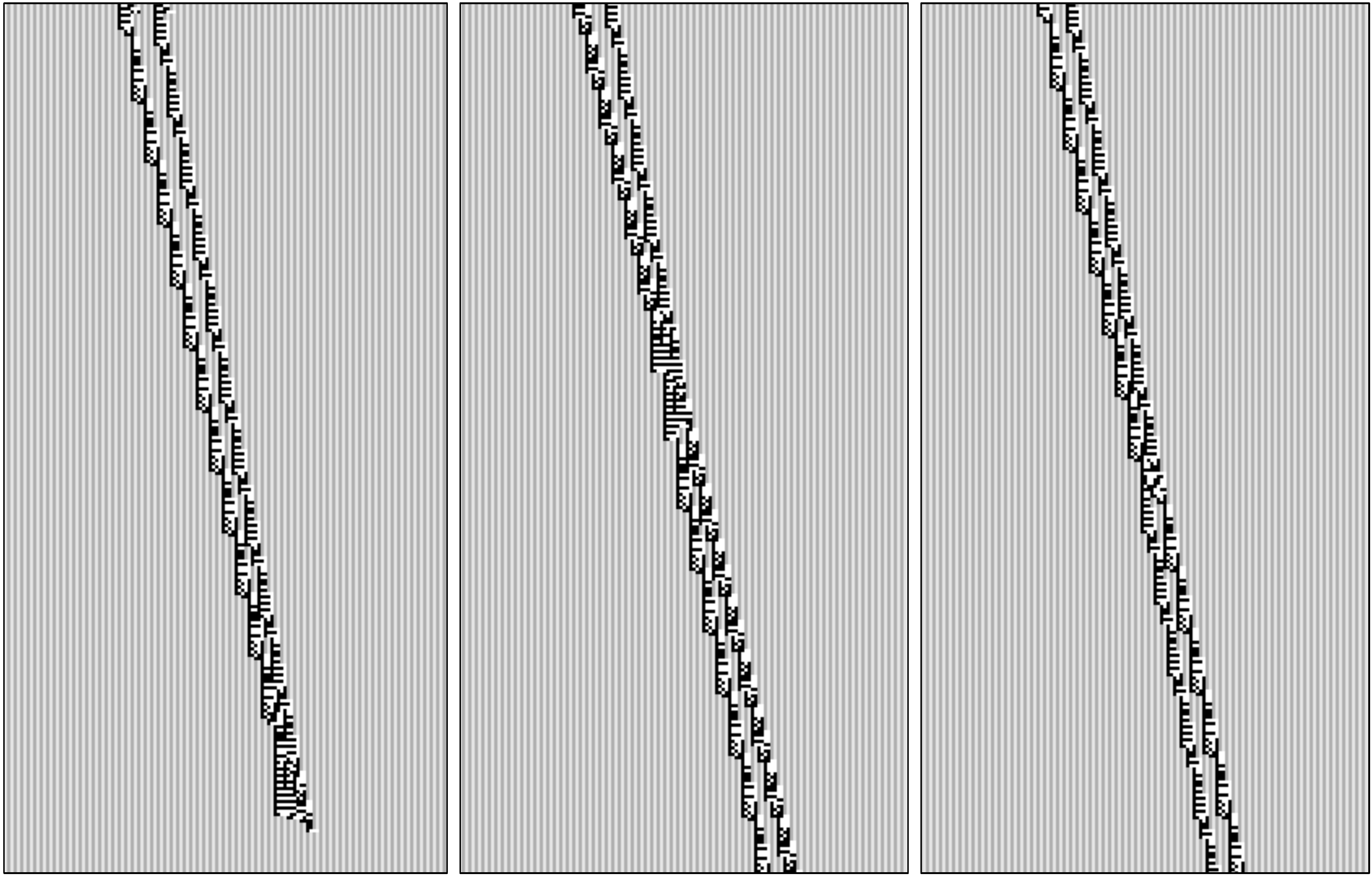}}
\caption{Some simple reactions display how to {\it delete} (left), {\it read} (centre) and {\it preserve} (rigth) information with gliders in Rule 30 with memory.}
\label{soliton}
\end{figure}

Some simple but interesting reactions from $G_{R30m}$ are illustrated in the fig.~\ref{soliton}. First reaction is annihilation of gliders $g_2$ and $g_1$. Second reaction is a transformation $g_3$ glider transforms $g_1$ glider to $g_2$. Third reaction is a soliton-like collision between gliders $g_2$ and $g_1$. The soliton reaction between gliders is particularly promising because it can be used to implement computation, e.g. as it has been done in {\it carry-ripple adder} embedded by phase coding solitons in parity CA~\cite{kn:PST86,kn:JSS01}.

\subsection{Quasi-chemistry of gliders}
Assuming gliders $g_1$, $g_2$ and $g_3$ are chemical species $a$, $b$ and $c$ in a well-stirred chemical reactor we can derive the following set of quasi-chemical reactions from the interactions between the gliders above:

\begin{equation}
\begin{array}{l}
a + b \stackrel{0.6}{\rightarrow} \epsilon\\
a + b \stackrel{0.2}{\rightarrow} 2a\\
a + b \stackrel{0.2}{\rightarrow} c\\
a + c \stackrel{0.44}{\rightarrow} b\\
a + c \stackrel{0.12}{\rightarrow} b+c\\
a + c \stackrel{0.22}{\rightarrow} 3a\\
a + c \stackrel{0.22}{\rightarrow} a\\
b + c \stackrel{0.5}{\rightarrow} b\\
b + c \stackrel{0.5}{\rightarrow} 2a + b
\end{array}
\label{reactions}
\end{equation}

\noindent where reaction rates are evaluated from the frequencies of the interactions.

\begin{figure}
\subfigure[]{\includegraphics[width=0.85\textwidth]{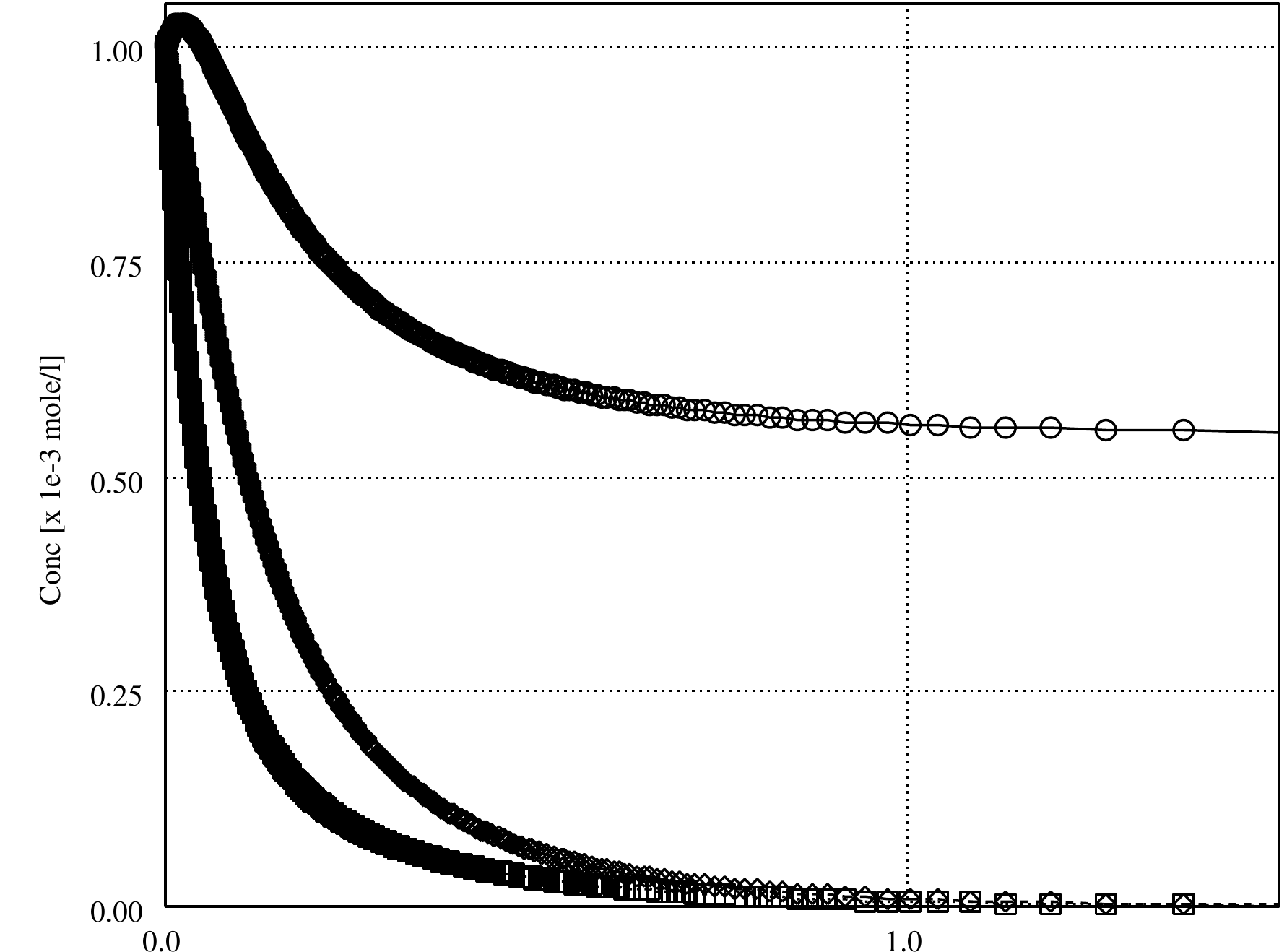}}
\subfigure[]{\includegraphics[width=0.85\textwidth]{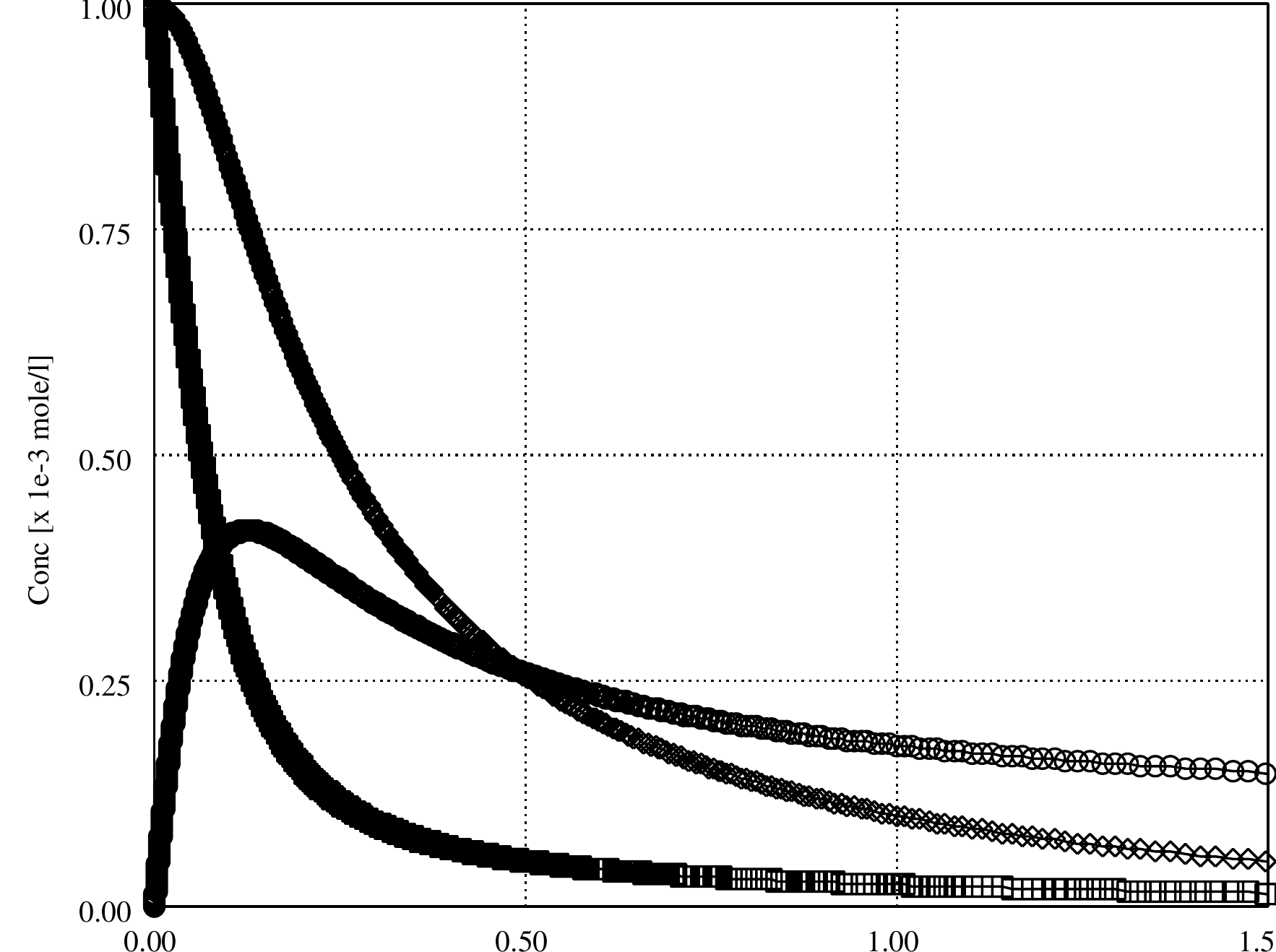}}
\caption{Dynamics of concentrations of species $a$ (circle), $b$ (diamond) and $c$ (square) governed by reaction eqs.~\ref{reactions}. (a)~Initial concentrations of all species are 0.001~mole/l. (b)~Initial concentrations of species $b$ and $c$ are 0.001~mole/l, species $a$ is nil. Horizontal axes is a time in $10^{-3}$~sec.}
\label{concentrations}
\end{figure}

We evaluated global dynamics of the quasi-chemical system (\ref{reactions}) with constant volume (which reflects finite size of the automaton lattice), constant temperature, and variable pressure using {\it Chemical Kinetics Simulator}.\footnote{\url{ http://www.almaden.ibm.com/st/computational_science/ck/?cks}} Fig.~\ref{concentrations} shows temporal dynamics of species concentrations in the system with $10^7$ molecules. 

In the initial conditions when all three species are present in equal concentrations (fig.~\ref{concentrations}a) we observe exponential decay of species $b$ and $c$ and stabilization of concentration of species $a$.  When only species $b$ and $c$ are present in a well-stirred reaction initially, they produce species $c$ in their reactions. This leads to outburst in species $a$ concentration (fig.~\ref{concentrations}b) on the background of exponential decline of species $c$ and $b$, until species $b$ and $c$ disappear and concentration of species $a$ becomes constant.

\subsection{Glider machines}
Experimentally found interactions between gliders can be simplified as follows, depending on distance $\sigma$ between interacting gliders:

$$
\begin{array}{llll}
\begin{array}{l}
\sigma = 3\\
b + a \rightarrow \{a,b \}\\
c + a \rightarrow b \\
c+ b \rightarrow b
\end{array} & 
\begin{array}{l}
\sigma = 4\\
b + a \rightarrow \{ \emptyset \}\\
c + a \rightarrow b \\
c+ b \rightarrow b
\end{array}
\begin{array}{l}
\sigma = 5\\
b + a \rightarrow \{ \emptyset \}\\
c + a \rightarrow \{ b, c \} \\
c+ b \rightarrow \{ a, b \}
\end{array} 
\begin{array}{l}
\sigma = 6\\
b + a \rightarrow  a\\
c + a \rightarrow \{ b, c \} \\
c+ b \rightarrow \{ a, b \}
\end{array} 
\end{array}
$$

Taking into account gliders' velocities (tab.\ref{tablaGlidersR30}), we can construct the following finite state indeterministic machine with an internal state $h$ and input state $p$, $h, p \in \{a, b, c, \emptyset \}$. The machine can be characterised by an input-output transition matrix $M=(m_{ij})_{i,j \in \{a, b, c, \emptyset \}}$, where for $j=h^t$, $i=p^t$, $m_{ij}=h^{t+1}$. The matrix has the following form:

$$
M = 
\begin{array}{l|llll}
h^{t+1} & a & b 									& c 						& \emptyset \\ \hline
a				& a &	\{ a, \emptyset \} 	& b							& a \\
b				& b	& b										& \{ a, b \}		& b \\
c				& c & c 									& c 						& c \\
\emptyset & \emptyset & \emptyset & \emptyset & \emptyset 
\end{array}
$$ 

Starting at a randomly chosen initial state  and subjected to random uniformly distributed input strings the machine will end in the state $c$ with probability $\frac{1}{4}$ and in the state $\empty$ set with probability $\frac{3}3{4}$. The machine starting in the initial $h$ generates the string $l(h)$ as follows: $l(\emptyset)=\emptyset^*$, $l(a)=(a b^* a^*)^* \emptyset^*$,$l(b)=(b^* a^*)^* \emptyset^*$, $l(c)=c^*$.

\section{Discussion}
We enriched elementary CA rule 30 with majority memory and demonstrated that by applying certain filtering procedures we can extract rich dynamics of travelling localizations, gliders, and infer sophisticated system of quasi-chemical reactions between the gliders. We shown that the majority memory increases nominal complexity but decreases statistical complexity of patterns generated by the CA. By applying methods of de Brujin diagrams and graph theory we proved that surjectivity of Rule 30 CA with memory and provide blue prints for future detailed analysis of glider dynamics.

\begin{figure}[th]
\centerline{\includegraphics[width=2.5in]{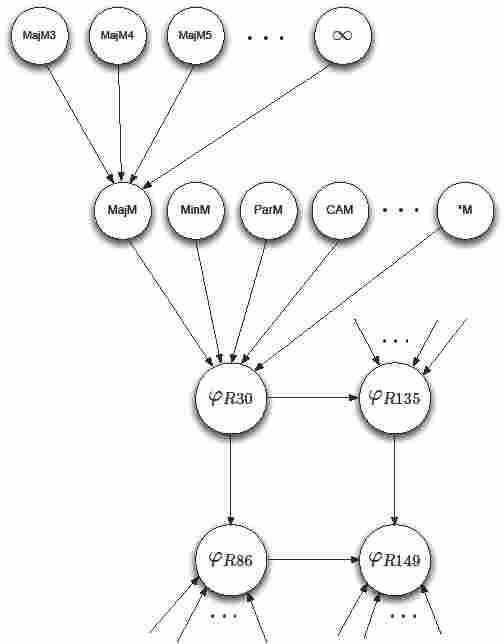}}
\caption{New family of elementary CA composed.}
\label{newClassECA}
\end{figure}

Recalling previous results on classification of one-dimensional CA~\cite{kn:WL92,kn:Wue94,kn:Wue99} we envisage that introduction of majority memory $\phi_{maj}$ into elementary CA will open a new field of research in selection of non-trivial rules of cell-state transitions and precise mechanics of relationships between chaotic and complex systems.

This is because Rule 30 was grouped into a cluster of rules with similar behaviour, that can be transformed one to another using combinations of reflexion, negation and complement (done by Wuensche in \cite{kn:WL92}). This way fig.~\ref{newClassECA} shows a diagram that explains how the original cluster for Rule 30 is presented in~\cite{kn:WL92}. Obviously the cluster can be arbitrarily enriched using not only $\phi_{maj}$ but any types of memory and $\tau$. Thus the dynamical complexity of automata with $\phi_{maj}$ is the same as for the set of function $\{\varphi_{R30}, \varphi_{R86}, \varphi_{R135}, \varphi_{R149}\}$, particularly because the local functions $\varphi_{R86}$ and $\varphi_{R149}$ are responsible for leftward motion of gliders.

Therefore, memory in elementary CA and others families of CA offer a new approach to discover complex dynamics based on gliders and non-trivial interactions between gliders. This can be substantiated by a number of different techniques, e.g. number-conservation~\cite{kn:BF02,kn:III04}, exhaustive search~\cite{kn:Epp02}, tiling~\cite{kn:MMS08,kn:Mar07}, de Bruijn diagrams~\cite{kn:MAM08}, $Z$-parameter~\cite{kn:Wue99}, genetic algorithms~\cite{kn:DMC94}, mean field theory~\cite{kn:Mc90} or from a differential equations point view~\cite{kn:Chu03}.

\section*{Acknowledgement}
Genaro J. Mart\'{\i}nez and Ramon Alonso-Sanz are supported by EPSRC (grants EP/F054343/1 and EP/E049281/1). J.~C. Seck-Touh-Mora is supported CONACYT (project CB-2007/083554).


\section*{Appendix A. Binary reactions and beyond}
We represent binary collisions in Rule 30, $\phi_{maj}$ and $\tau=8$  in the form $g_j \xrightarrow{space} g_i$, where $j > i$, $g \in G_{R30}$ and {\it space} is the interval between gliders given by the number of strings $w_b$. Collisions producing $\epsilon$ mean annihilation of gliders. Reactions are developed using increasing the distance between gliders before collision. \\

\noindent {\bf Collisions of type $g_2 \rightarrow g_1$}

\begin{enumerate}
\item $g_2 \xrightarrow{3} g_1 = g_1 + g_2$ (soliton)
\item $g_2 \xrightarrow{4} g_1 = \epsilon$
\item $g_2 \xrightarrow{5} g_1 = \epsilon$
\item $g_2 \xrightarrow{6} g_1 = 2g_1$
\item $g_2 \xrightarrow{7} g_1 = \epsilon$
\item $g_2 \xrightarrow{8} g_1 = g_3$
\end{enumerate}

\noindent {\bf Collisions of type $g_3 \rightarrow g_1$}

\begin{enumerate}
\item $g_3 \xrightarrow{3} g_1 = g_2$
\item $g_3 \xrightarrow{4} g_1 = g_2$
\item $g_3 \xrightarrow{5} g_1 = g_2 + g_3$
\item $g_3 \xrightarrow{6} g_1 = g_1^3$
\item $g_3 \xrightarrow{7} g_1 = g_1 + g_1^2$
\item $g_3 \xrightarrow{8} g_1 = g_2$
\item $g_3 \xrightarrow{9} g_1 = g_1$
\item $g_3 \xrightarrow{10} g_1 = g_2$
\item $g_3 \xrightarrow{11} g_1 = g_1$
\end{enumerate}

\noindent {\bf Collisions of type $g_3 \rightarrow g_2$}

\begin{enumerate}
\item $g_3 \xrightarrow{3} g_2 = g_2$
\item $g_3 \xrightarrow{4} g_2 = g_2$
\item $g_3 \xrightarrow{5} g_2 = 2g_1 + g_2$
\item $g_3 \xrightarrow{6} g_2 = 2g_1 + g_2$
\end{enumerate}

\noindent {\bf Some other reactions with packages of gliders}

\begin{enumerate}
\item $g_2 \xrightarrow{5} 2g_1 = g_1$
\item $g_2 \xrightarrow{8} 2g_1 = 3g_1$
\item $g_2 \xrightarrow{10} 2g_1 = g_2$ (sequence $g_2$, $g_3$, $g_2$)
\item $2g_2 \xrightarrow{6} g_1 = 3g_1$
\item $2g_2 \xrightarrow{7} g_1 = g_2$
\item $2g_2 \xrightarrow{8} g_1 = g_1 + g_3$
\item $g_3 \xrightarrow{5} 2g_1 = \epsilon$
\item $g_3 \xrightarrow{6} 2g_1 = g_3$
\item $g_3 \xrightarrow{7} 2g_1 = g_1 + g_2$
\item $g_3 \xrightarrow{8} 2g_1 = g_1^3 + g_1$
\item $g_3 \xrightarrow{9} 2g_1 = g_1^4$
\item $g_3 \xrightarrow{10} 2g_1 = 2g_1$
\item $2g_3 \xrightarrow{5} g_2 = g_2$ (wall)
\end{enumerate}

\section*{Appendix B. Coding gliders $G_{R30m}$}
We can enumerate strings conforming gliders in Rule 30 with $\phi_{maj}$ and $\tau=8$, in given initial conditions and using ``phases'' (we omit strings that do not produce gliders).

Note that such strings evolve initially with $\varphi_{R30}$ and a value of $\tau$ given, then $\phi_{maj}$ will open these strings when memory works. Thus we can code initial conditions with gliders in CA with memory, that also was implemented in OSXLCAU21 system to get our simulations.

\begin{table}[th]
\centering
\begin{tabular}{|l|l|l|}
\hline
$g_1$ glider & $g_2$ glider & $g_3$ glider \\
\hline
1 -- $g_1$ & 100 -- $g_2$ & 110110 -- $2g_1$ join \\
111 -- $g_1$ & 111 -- $g_1$ & 101110 -- $g_3$ \\
10000 -- $g_1$ & 10000 -- $g_1$ & 111100 -- $2g_1$ \\
11110 -- $g_1$ & 11110 -- $g_1$ & 100011 -- $2g_1$ join \\
100 -- $g_2$ & 1100110 -- $g_2 + g_3$ & 1011 -- $g_3$ \\
11110 -- $g_1$ & 1010110 -- $g_1$ & \\
10000 -- $g_1$ & 1101100 -- $g_2$ & \\
11001 -- $g_2 + g_3$ & & \\
\hline
\end{tabular}
\caption{Strings evolving in gliders of $G_{R30m}$.}
\label{stringGlider}
\end{table}

Tab.~\ref{stringGlider} enumerates strings for each glider represented as a tiling so we know their ``phases''~\cite{kn:MMS06,kn:MMS08}. Although in this case, it was difficult to classify such strings as regular expressions because not all strings from the tiling representation evolve into gliders.

\end{document}